# Serverless computing provides on-demand high performance computing for biomedical research

Dimitar Kumanov[1,#], Ling-Hong Hung[1,#], Wes Lloyd[1], Ka Yee Yeung[1,*]

[1]Institute of Technology, University of Washington, Tacoma, WA 98402, USA

[#]Equal contributions

[*]Corresponding author

**ABSTRACT**

Cloud computing offers on-demand, scalable computing and storage, and has become an essential resource for the analyses of big biomedical data. The usual approach to cloud computing requires users to reserve and provision virtual servers. An emerging alternative is to have the provider allocate machine resources dynamically. This type of 'serverless' computing has tremendous potential for biomedical research in terms of ease-of-use, instantaneous scalability and cost effectiveness. In our proof of concept example, we demonstrate how serverless computing provides low cost access to hundreds of CPUs, on demand, with little or no setup. In particular, we illustrate that the all-against-all pairwise comparison among all unique human proteins can be accomplished in approximately 2 minutes, at a cost of less than $1, using Amazon Web Services Lambda. This is a 250x speedup compared to running the same task on a typical laptop computer.

Cloud computing has become an essential resource in the analyses of big biomedical data by offering massive scalable computing and storage, data sharing and on-demand access to resources and applications [1, 2]. Most major public cloud vendors offer tools for biomedical research, including Google Cloud Genomics [3], Amazon Web Services (AWS) [4] and Microsoft Genomics [5]. Currently, bioinformatics use cases benefit from leveraging cloud computing resources by provisioning on-demand computing infrastructure [6, 7], providing secure access to protected data and scalable software platforms [8, 9]. Recently, serverless computing, also referred to as the Function-as-a-Service paradigm, has emerged as an evolved simplified programming model that can be used to create scalable cloud applications with reduced configuration and management overhead [10]. Instead of the user reserving and provisioning a set of virtual servers, the provider allocates the machine resources as needed. Specifically, serverless computing enables access to hundreds of CPUs on demand, with little or no setup. Here, we demonstrate the power and economy of scale of a serverless computing approach to simultaneously allocate hundreds of CPUs to complete a computationally intensive bioinformatics task in approximately 2 minutes, providing more than a 250x speedup, at a cost of less than $1. In contrast, running the same task on a typical laptop computer required 8.7 hours.

A major barrier to widespread adoption of the cloud for large scale parallel bioinformatics jobs is the need to provision and configure virtual servers before computation can proceed. Setup requires expertise and can consume more time than the actual compute. In contrast, code snippets can be easily deployed to the cloud in the form of microservices using serverless computing with resource provisioning, monitoring, scalability and fault tolerance provided automatically [11, 12]. In addition to the ease of use, these microservices can be launched in under a second, require no provisioning or specification of instances and are inexpensive. This is a true on-demand model: there is no cost to the user when the user's function code is not used. This is in contrast to cloud models where the user is charged for the cloud infrastructure even when virtual servers are idle. Most major public cloud vendors provide serverless computing capabilities, including the AWS Lambda [13], Google Cloud Functions [14], Microsoft Azure Functions [15] and IBM Cloud Functions [16].

Here, we present a proof-of-concept case study performing sequence alignment comparing 20,000 protein sequences using serverless computing. Our case study demonstrates tremendous potential for leveraging serverless computing computational resources for biomedical research in terms of ease-of-use, instantaneous scalability and cost effectiveness. While there are major advantages of serverless computing (i.e. users pay only for resources used when serverless functions are running), cloud providers place limits on the available computational resource for hosting serverless code including: the maximum number of concurrent client requests and the maximum memory and CPU resources available for each function invocation [11]. On AWS Lambda, functions execute with the following constraints: 500MB of disk space, 5 minutes maximum runtime, 3GB RAM, and 2 vCPUs. By default, Lambda functions are limited to 1,000 concurrent client requests per account, though this limit can easily be increased on request. We show that these constraints can be addressed using distributed approaches similar to those used in algorithms that leverage graphics processing units (GPUs).

Smith-Waterman [17] is a dynamic programming algorithm for comparing protein sequences. Performing similarity comparisons of protein sequences is a common precursor to clustering into related families or functional groups. The Similarity Matrix of Proteins (SIMAP) project applies Smith-Waterman to compare all-against-all pairs of known protein sequences covering all major protein public databases, a task that is expected to take years to compute [18]. For our demonstration, we perform the all-against-all pairwise comparison with human proteins using the Striped Smith-Waterman (SSW) implementation [19]. Our input data consist of 20,336 unique human protein sequences from UniProt [20]. We partition these 20,336 sequences into 41 subsets, each consisting of approximately 500 sequences stored in Amazon's Simple Storage Service (S3). The calculation of all the pairwise interactions between sequences in a pair of subsets (i.e. 500 vs. 500 proteins) is small enough to be run by individual Lambda functions with 1.5 GB memory capacity. This results in a total of 861 unique tasks to compare each of these 41 subsets to another subset including itself. To run a comparison task using AWS Lambda, the user bundles a script with the required dependencies and uploads them. The function executes on Lambda using Amazon Linux [21]. To profile the performance of running Smith-Waterman on AWS Lambda, we deployed a scheduler script written in Python on an Elastic Compute Cloud (EC2) instance, and on a personal laptop to coordinate the execution of all 861 tasks concurrently. This script acts as the client to invoke AWS Lambda independently for each respective task utilizing the Striped Smith-Waterman (SSW) executable. Our Lambda function writes similarity scores to a file on S3, and reports completion by sending a message to the AWS Simple Queue Service (SQS). The client Python script monitors SQS for messages to determine when the whole job has completed all 861 tasks. Our code and documentation are publicly available on GitHub. (https://github.com/BioDepot/TaskPerform_AWSLambda). An overview of our approach is summarized in **Figure 1**.

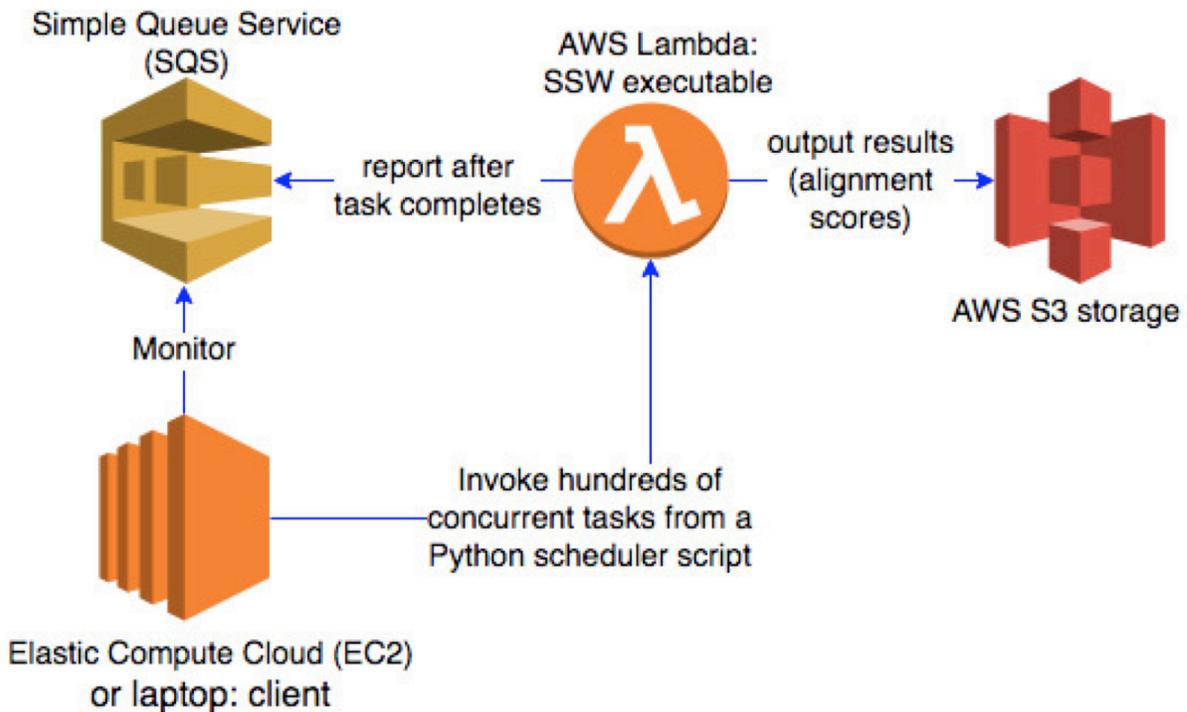

**Figure 1:** Overview of our approach. The Python scheduler script running on an Elastic Compute Cloud (EC2) instance or a personal laptop invokes hundreds of concurrent tasks on AWS lambda in parallel. In our use case study, a Python scheduler script reads two subsets of proteins (each subset consists of approximately 500 protein sequences), calls the Striped Smith-Waterman (SSW) executable, writes the similarity scores to a file on S3, and reports completion by sending a message to the AWS Simple Queue Service (SQS), a message queuing service. The EC2 instance or the laptop monitors SQS to determine when the whole job has completed.

When the Lambda functions are invoked by the Python script running on a laptop computer, the average execution time is 2.20 minutes. This is in contrast to a run time of 8.7 hours when performing all the tasks on the laptop without using AWS Lambda. Similarly, the average execution time is just 1.28 minutes when the Lambda functions are invoked by the Python script running on an AWS EC2 instance. The estimated hosting cost using AWS Lambda for this application for either client is under $1. Please refer to the Methods section for details of our empirical experiments. Table 1 shows details of our cost analyses.

**Table 1:** Comparison of average execution time and estimated cloud computing costs. Invoking AWS Lambda functions with 1.5GB memory capacity from a Python scheduler script on an EC2 instance (m5.x24.xlarge) or on a personal laptop takes approximately 2 minutes at the cost of under $1. On the contrary, running this same protein sequence alignment task takes 8.7 hours (522 minutes) on the laptop without using AWS Lambda.

|  | EC2 client + AWS Lambda | Laptop client + AWS Lambda | laptop without AWS Lambda |
| --- | --- | --- | --- |
| average execution time (minutes) | 1.28 | 2.20 | 522 |
| Average total AWS Lambda Duration across concurrent tasks (CPU minutes) | 542.14 | 530.69 | |
| Total AWS Lambda Duration Cost ($) | $0.83289 | $0.81531 | |
| Total AWS Lambda Invocation (Requests) Cost ($) | $0.00017 | $0.00017 | |
| Total SQS Queue Cost ($) | $0.00103 | $0.00103 | |
| Cost of EC2 instance ($1.6 an hour using a spot instance) | $0.03413 | $0.00000 | |
| Total estimated AWS costs ($) | $0.86823 | $0.81651 | |

Our example demonstrates how a computationally intensive bioinformatics task can exploit serverless cloud computing to dramatically speed up execution time at low cost. This approach can be applied to other problems that can be partitioned into small sub-problems. This is commonly done for example, in GPU implementations, where a large task is divided into sub-tasks that are operated on by small groups of GPU compute units. Sequence alignment, protein-folding, and deep-learning are computationally intensive bioinformatics tasks with GPU implementations [22-24] where a serverless computing approach may prove useful. With the ability of serverless cloud computing to quickly leverage hundreds of CPUs, computational power that was once the exclusive domain of supercomputers is now easy to access, available on demand, and at low cost, to help solve resource intensive bioinformatics problems.

## METHODS

*EC2 client + AWS Lambda.* We repeated our empirical experiments invoking AWS Lambda functions with 1.5GB memory capacity from an EC2 client (m5.x24.xlarge) 10 times. The observed execution time completing all 861 concurrent Lambda functions ranges from 59.003 seconds to 114.804 seconds, with a mean of 76.647 seconds (=1.28 minutes) and a median of 62.286 seconds (=1.04 minutes).

*Laptop client + AWS Lambda.* We repeated our empirical experiments invoking AWS Lambda functions with 1.5GB memory capacity from our laptop client (Intel i5-7200U Cpu @ 2.50 GHz 2.70 GHz processor, 8 GB of RAM) 3 times. The observed execution time completing all 861 concurrent Lambda functions ranges from 127.808 seconds to 139.953 seconds, with a mean of 132.199 seconds (=2.20 minutes) and a median of 128.835 seconds (=2.15 minutes).

*Laptop without AWS Lambda.* Our empirical experiments are performed using a laptop computing with Intel i5-7200U CPU @ 2.50 GHz 2.70 GHz processor, 8 GB of RAM running Ubuntu16.04. The experiments are performed using Oracle VM Virtual Box with the following settings: 4096 MB RAM and 1 core processor at 100% execution cap.

## ACKNOWLEDGEMENTS

We would like to acknowledge support from the AWS Cloud Credits for Research (to Lloyd and Yeung) for providing cloud computing resources.

## Funding

This work is funded by NIH grant R01GM126019.

## Availability of data and materials

- Project name: Using AWS Lambda for bioinformatics applications
- Project home page: https://github.com/BioDepot/TaskPerform_AWSLambda
- Operating system(s): Platform independent
- Programming language: Python
- Other requirements: Amazon Web Services (AWS) Lambda
- License: CC BY-NC-ND
- Any restrictions to use by non-academics: none

## Authors' contributions

D.K. wrote all the code, performed the empirical experiments, collected results from the empirical studies and created Figure 1. L.H.H. designed the case study and helped to write the manuscript. W.L. contributed ideas on serverless computing and helped to write the manuscript. K.Y.Y. drafted the manuscript, helped with analyzing results from the empirical studies and directed the study. L.H.H., W.L. and K.Y.Y. all worked closely with D.K.

## Ethics approval and consent to participate

Ethics approval is not applicable to this work.

## Competing interests

None.